\documentclass{ws-procs9x6}
\begin{document}
\vspace*{-5mm}
\title{PHOTO- AND ELECTROPRODUCTION OF KAONS}

\author{P. BYD\v{Z}OVSK\'Y}

\address{Nuclear Physics Institute, 
25068 \v{R}e\v{z} near Prague, Czech Republic}

\bodymatter

\begin{abstract}
Isobar models for the electromagnetic production of K$^+$ and K$^0$ 
are discussed with emphasis on the K$^0$ photoproduction off neutron. 
Predictions of the models for the K$^0$ photoproduction on deuteron 
are compared with the recent data and some properties of the 
elementary amplitudes are shown.   
\end{abstract}

\section{Introduction}

The electromagnetic production of strangeness on nucleon provides us 
with additional information about the structure and interactions of 
baryons. Beyond investigation of the reaction mechanism, form factors 
of hadrons and new ``missing'' resonances one also needs a correct 
description of the process for a study of more complex processes such 
as the electroproduction of hypernuclei. A good quality description 
of the elementary process minimises uncertainty in calculations 
of the excited spectra for the hypernucleus electroproduction~\cite{ProdH}.
Many new good quality experimental data for the K$^+$ production 
collected in JLab, ELSA, SPring-8, and GRAAL allow performing a 
thorough analysis of the elementary process. Moreover, the recent 
data on the photoproduction of K$^0$ off deuteron from Tohoku 
University~\cite{Tsu08} facilitate doing more rigorous tests of 
elementary amplitudes~\cite{K0d04,Tsu08}.

The electroproduction process can be formally reduced to investigation 
of the binary process of the photoproduction by virtual photons since 
the electromagnetic coupling constant is small enough to justify 
the one-photon approximation. 
There are several approaches to treat the photoproduction process. 
Among them the isobar models based on the effective Lagrangian 
description considering only the hadronic degrees of freedom are 
suitable for their further use in the more complex calculations. 
Other approaches are eligible either for higher energies 
($E_{\gamma}>4$ GeV), the Regge model~\cite{Regge}, or to the 
threshold region, the Chiral Perturbation Theory~\cite{ChPT}. 
Quark models~\cite{SagQM} are too complicated for their further 
use in the hypernuclear calculations. Another approach, aimed 
at the forward-angle production, is the hybrid Regge-plus-resonance 
model~\cite{RPR} in which the background part of amplitude is 
generated by the $t$-channel Regge-trajectory exchange and 
the resonant behaviour is shaped by the s-channel resonances 
like in an isobar model. 

\section{Isobar Models}\label{prdN} 
 
In the effective hadron Lagrangian approach, various channels 
connected via the final-state interaction (the meson-baryon 
rescattering processes) have to be treated simultaneously to take 
unitarity properly into account~\cite{CouCh,Chi01}.  
In Ref.~\refcite{Chi01} the coupled-channel approach has been 
used to include effects of the $\pi$N intermediate states in 
the p($\gamma$,K$^+$)$\Lambda$ process. However, tremendous 
simplifications originate in neglecting the rescattering effects 
in the formalism assuming that they are included to some extent 
by means of effective values of the strong coupling constants 
fitted to data. This simplifying assumption was adopted in many 
of the isobar models, e.g., Saclay-Lyon (SLA)~\cite{SLA}, 
Kaon-MAID (KM)~\cite{Ben99}, and Janssen {\it et al.}~\cite{Jan01}.

In isobar models, the amplitude obtains contributions from 
the Born terms and exchanges of resonances. Due to absence of 
a dominant resonance in the process (unlike in the $\pi$ and 
$\eta$ photoproduction) large number of possible combinations 
of the resonances with mass smaller than 2 GeV must be taken 
into account~\cite{SLA}. This number of models is limited 
considering constraints set by SU(3)~\cite{SLA,Ben99} and  
crossing symmetries~\cite{WJC92,SLA} and by duality hypothesis~\cite{WJC92}. 
Adopting the SU(3) constraints to the two main coupling constants, 
however, makes the contribution of the Born terms nonphysically 
large~\cite{Jan01}. To reduce this contribution, either hyperon 
resonances~\cite{SLA} or hadron form factors~\cite{Ben99} must 
be added, or a combination of both\cite{Jan01}. The hadron form 
factors which simulate a structure in the strong vertex are included 
in the KM and Janssen models maintaining the gauge invariance of 
amplitude~\cite{Ben99,Jan01}. Here we use the KM and SLA models. 

The strong coupling constants in the K$^0\Lambda$ and K$^+\Lambda$ 
channels are related via the SU(2) isospin symmetry: 
$g_{{\rm K}^+\Lambda {\rm p}}= g_{{\rm K}^0\Lambda {\rm n}}$ 
and $g_{{\rm K}^+\Sigma^0 {\rm p}}= -g_{{\rm K}^0\Sigma^0 {\rm n}}$. 
In the electromagnetic vertexes a ratio of the neutral to charged 
coupling constants have to be known. For the nucleon and its resonances 
the ratio can be related to the known helicity amplitudes of the 
nucleons~\cite{PDP} whereas for the kaon resonances the ratio relates 
to the decay widths known only for the K$^*$ meson~\cite{PDP}: 
r$_{\rm K^*}=-\sqrt{\Gamma_{{\rm K}^{*0}\rightarrow {\rm K}^0\gamma}/ 
\Gamma_{{\rm K}^{*+}\rightarrow {\rm K}^+\gamma}}=-1.53$ where the sign 
was set from the quark model prediction. Since the decay widths of 
the K$_1$ meson are unknown the appropriate ratio,  r$_{\rm K_1}$, 
have to be fixed in the models. It was fitted to the K$^0\Sigma^+$ 
data in KM~\cite{Ben99}, r$_{\rm K_1}=-0.45$, but it is a free 
parameter in SLA (see also Refs.~\refcite{Tsu08} and \refcite{K0d04}). 
For hyperons, the electromagnetic vertex is not changed in the 
K$^0\Lambda$ channel.

\section{Photoproduction on the Deuteron}

The inclusive cross section for the d($\gamma$,K$^0$)YN' or 
d($\gamma$,$\Lambda$)KN' process (Y stands for $\Lambda$ or $\Sigma$ 
and N' for proton or neutron) in the threshold region is calculated 
in the impulse approximation in which the nucleon N' acts as 
a spectator~\cite{K0d04}. Contributions of the final-state 
interaction (FSI) to the inclusive cross section were shown to be 
small in the studied kinematical region~\cite{Agus}.  
Moreover, we assume that a part of the KY rescattering effects 
is absorbed in the coupling constants of the elementary amplitude 
and that the KN interaction is weak on the hadronic scale. 
The main lack of precision, therefore, comes from ignoring 
the YN FSI. 

In the calculations we have chosen the {\it off-shell} 
approximation~\cite{K0d04} in which the four-momentum is conserved in 
the production vertex forcing the target nucleon off its mass shell. 
The momentum distribution of the target nucleon is described by the 
nonrelativistic Bonn deuteron wave function OBEPQ~\cite{Bonn}.
More details about the calculations can be found in Ref.~\refcite{K0d04}.

\section{Results and Discussion}

The existing isobar models do not give satisfactory description 
of the new data on the K$^+$ photoproduction, especially of the 
spin observables~\cite{Bra07}. Here we mention only the difference 
in results of the models for the forward kaon-angle cross section  
which causes substantial uncertainty in predictions for the 
hypernucleus photoproduction cross sections~\cite{ProdH}. 
At very forward angle, $\theta_{\rm K}^{\rm lab}\!= 3^o$, and photon lab 
energy  1.42 GeV, the lab cross sections predicted by the SLA and KM 
models are 2.24 and 1.60 $\mu$b/sr, respectively. This 30\% discrepancy 
gives corresponding difference in results for the photoproduction 
of $^{12}$B$_\Lambda$ at 1.42 GeV:   
the excitation cross sections for the two main multiplets ($\Lambda$ 
in $s$ and $p$ state is coupled to the ground state of the core 
nucleus $^{11}$B)  
are 169 and 173 nb/sr for SLA and 119 and 127 nb/sr for KM, respectively. 
We see that the elementary amplitude needs to be better fixed 
at forward angles to give reliable results for the hypernucleus 
calculations. Studying the process in other channels, e.g.,  
n($\gamma$,K$^0$)$\Lambda$, might help in this respect.
 
%
%
\begin{figure}[th]
\centerline{\epsfxsize=4.2cm\rotatebox{-90}
         {\epsfbox{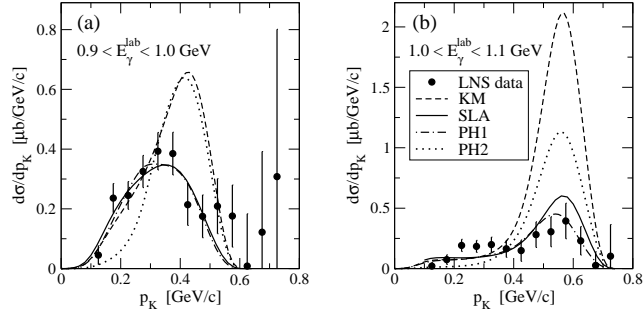}}}   
\caption{Inclusive energy-averaged, (a) 0.9 $<$ E $<$ 1.0 GeV and (b) 
 1.0 $<$ E $<$ 1.1 GeV, and kaon-angle-integrated,  
0.9 $<$ cos($\theta_K$) $<$ 1.0, momentum spectra for the 
d($\gamma$,K$^0$)YN' process. Calculations using various elementary 
amplitudes are performed for the $\Lambda$p final state. Data are 
from Ref.~\refcite{Tsu08}. \label{figure1}}
\end{figure}
%
%
\begin{figure}[bh]
\centerline{\epsfxsize=4.2cm\rotatebox{-90}
           {\epsfbox{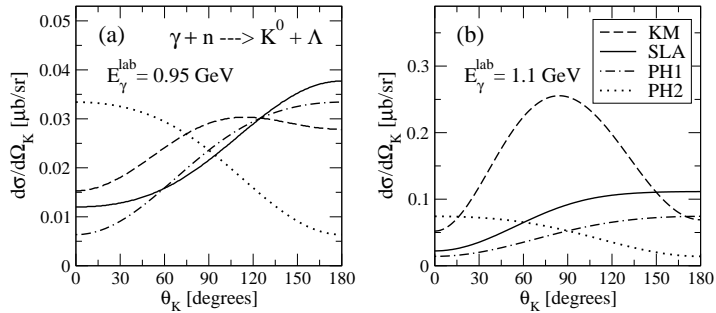}}}   
\caption{Predictions of the models for the c.m. differential cross section 
in the photo-production of K$^0$ on neutron are shown for the photon lab 
energies 0.95 and 1.1 GeV.\label{figure2}}
\end{figure}
In Figure~\ref{figure1} results for the inclusive momentum spectra 
of the d($\gamma$,K$^0$)$\Lambda$p process calculated with different 
elementary amplitudes are shown in comparison with the recent data 
from LNS of Tohoku University~\cite{Tsu08}. Contributions of the $\Sigma$ 
channels are very small in the lower-energy region (l.e.r.) 
(Fig.~\ref{figure1}a) and at the spectator-kinematics region (the main 
peak) in Fig.~\ref{figure1}b~\cite{Tsu08}. The KM model do not describe 
the data well, especially at the spectator-kinematics region. Results 
of KM cannot be improved even if the r$_{\rm K_1}$ parameter is fitted 
to the l.e.r. data: $\chi^2_{\rm n.d.f.}$= 3.64 with r$_{\rm K_1}$= -2.34. 
The SLA model with r$_{\rm K_1}$= -2.09, fitted to the l.e.r. data 
($\chi^2_{\rm n.d.f.}$= 0.88)~\cite{Tsu08}, gives good results in both 
energy regions. The angular dependence of the elementary cross sections 
differs for these models as shown in Fig.~\ref{figure2}. 
To point out importance of the angular dependence of cross section 
the phenomenological prescriptions PH1 and PH2 
(see Ref.~\refcite{Tsu08} for details) were used. PH1 which possesses 
the backward-peaked cross section, Fig.~\ref{figure2}, gives much better 
results for d($\gamma$,K$^0$)$\Lambda$p than PH2 with the opposite 
angular dependence (Fig.~\ref{figure2}). This comparison shows 
that the Tohoku data prefer models which give a backward-peaked 
cross section for the n($\gamma$,K$^0$)$\Lambda$ reaction.

%
%
\begin{figure}[t!]
\centerline{\epsfxsize=4.2cm\rotatebox{-90}
           {\epsfbox{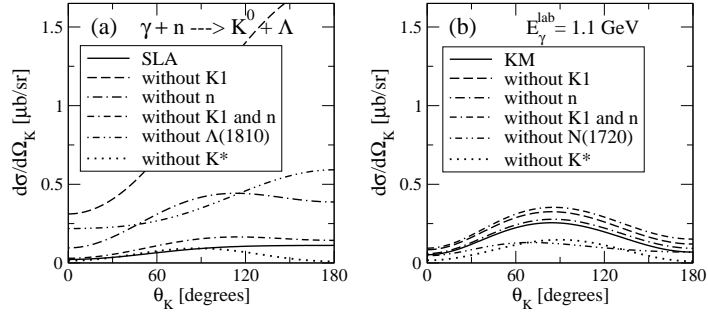}}}   
\caption{Angular dependence of the c.m. cross section for the photoproduction 
of K$^0$ on neutron at 1.1 GeV. Results for the Saclay-Lyon (a) and 
Kaon-MAID (b) models are shown when some of the contributions are switched off. 
\label{figure3}}
\end{figure}
Significance of various contributions in the elementary cross section at 
$E_\gamma = 1.1$ GeV is shown for SLA and KM in Fig.~\ref{figure3}. 
If the contribution of neutron or K$_1$ is omitted in SLA the cross 
sections change largely, Fig.~\ref{figure3}a, whereas the omission of 
both terms simultaneously makes a very small effect. This shows that 
the main role of K$_1$, with the fitted parameter r$_{\rm K_1}$, 
is to compensate the neutron contribution. This phenomenon  
is absent in KM, Fig.~\ref{figure3}b, since the neutron contribution  
itself is already strongly suppressed by the hadron form factor. 
In both models the contribution of K$^*$ is important at large 
$\theta_{\rm K}$. The hyperon resonances, included in SLA but not 
in KM, also contribute notably at backward angles (Fig.~\ref{figure3}a). 
The nucleon resonances give a negligible contribution  
at $\theta_{\rm K}\approx 180^\circ$ in both models.  

%
%
\begin{figure}[b!]
\centerline{\epsfxsize=4.2cm\rotatebox{-90}
          {\epsfbox{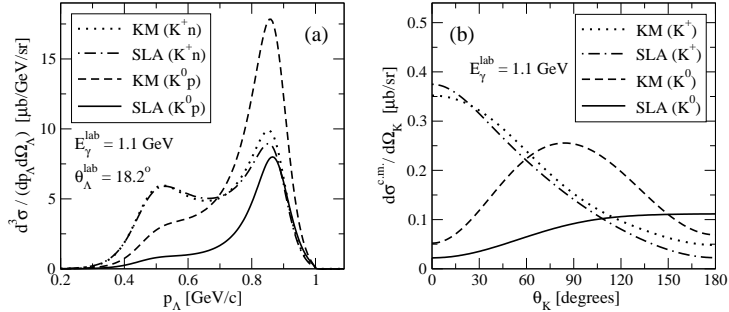}}}   
\caption{Predictions of the Saclay-Lyon (SLA) and Kaon-MAID (KM) 
models for (a) momentum spectra of the $\Lambda$ photoproduction 
on deuteron with K$^+$n and K$^0$p final states; (b) angular dependence  
of the c.m. cross section for the $\Lambda$ photoproduction on the proton 
and neutron with associated K$^+$ and K$^0$ in the final state, respectively. 
\label{figure4} }
\end{figure}
In Figure~\ref{figure4} predictions of the KM and SLA models for 
the momentum spectra of d($\gamma$,$\Lambda$)K$^+$n and 
d($\gamma$,$\Lambda$)K$^0$p (a) are shown with the angular dependence 
of the elementary cross sections (b). The models give significantly 
different results for the K$^0$ production whereas they accord for 
the K$^+$ channel. This shows that the K$^0$ photoproduction on 
deuteron provides with additional information about dynamics of the 
elementary process. 
 
To summarise: the isobar models for the K$^+$ photoproduction need 
revision at forward kaon angles before they can give 
reliable predictions for the hypernuclear calculations.
The data on the d($\gamma$,K$^0$)YN' reaction near the threshold 
clearly prefer the models for the n($\gamma$,K$^0$)$\Lambda$ reaction 
which give enhancement of the cross section at the backward kaon angles. 
These data can, therefore, discriminate between the models which 
otherwise fit the p($\gamma$,K$^+$)$\Lambda$ data equally well. 
\vspace*{2mm}

\noindent {\small The author wishes to thank the organisers 
for their kind invitation to this highly stimulating Symposium. 
This work was supported by the Grant Agency of the Czech Republic, 
grant 202/08/0984.}

\end{document}